\documentstyle[prd,aps]{revtex}

\def\L{{\cal L}}
\def\s{\sigma}
\def\l{\label}
\def\p{\phi}
\def\d{\delta}
\def\ts{\textstyle}

\def\beq{\begin{equation}}
\def\eeq{\end{equation}}
\def\bea{\begin{eqnarray}}
\def\eea{\end{eqnarray}}

\def\ov{\overline}
\def\arr{\leftrightarrow}
\def\sp{\sigma({\cal L})}
\def\csp{\overline{\sigma({\cal L})}}
\def\lam{\lambda}
\def\ts{\textstyle}

\begin{document}
\input epsf
\draft
\renewcommand{\topfraction}{0.8}
\twocolumn[\hsize\textwidth\columnwidth\hsize\csname
@twocolumnfalse\endcsname
\preprint{}
\title { \bf Inflationary reheating classes via 
spectral methods} 
\author{Bruce A. Bassett \footnote{Email:bruce@stardust.sissa.it}}
\address{ International School for Advanced Studies, Via Beirut 2-4, 
34014, Trieste, Italy}
\date{\today}
\maketitle
\begin{abstract}
Inflationary reheating is almost completely controlled by the Floquet 
indices, $\mu_k$. Using spectral theory we  demonstrate that the 
stability bands (where  $\mu_k = 0$) of the  Mathieu and Lam\'e equations 
are destroyed even  in Minkowski spacetime,  leaving a fractal Cantor 
set  or a measure zero  set of stable modes in the cases where the  
inflaton evolves in an  almost-periodic or stochastic manner 
respectively. These two types of potential model the expected  
multi-field and quantum  backreaction effects during reheating.  
\end{abstract} 
\vskip2pc]

\section {Introduction}

Inflation is perhaps the heir apparent in high energy cosmology. No other
theory can be cast successfully in so many guises. Yet,  to be successful,
any inflationary scenario must reheat the universe. This 
epoch is typically very short and has become a powerful virtual laboratory 
for researching non-equilibrium quantum field theory in curved spacetime 
\cite{KLS94,TYB94,KLS95,Boyan,Yosh95,KT96,PR97,GPR97,KLS97,GKLS97,BL97}. 
It is defined  by rapidly evolving
paradigms (see the left half of Fig. 1), attempting to deal with the
non-perturbative  processes occurring in  preheating \cite{KLS94}. 
The defining characteristic of a preheating model are its Floquet
indices $\mu_k$, which control  the number of  produced particles $n_k 
\sim \int dk k^2 e^{2 m \mu_k t}$ and variances of 
the fields \cite{KLS97}.

Of particular interest are the issues of how the $\mu_k$ vary as 
(i) the parameters  (typically the coupling constants) of the theory 
are changed, and (ii) the functional form of the inflaton evolution is 
altered. The first has been studied in depth in the case where the 
inflaton evolution is exactly periodic  and is the very origin 
of preheating. The second issue is much less studied 
and is  the subject of this paper where we introduce spectral methods 
as an elegant  classification tool for inflationary reheating classes; 
though it also applies to the parametric amplification of  
gravitational waves \cite{BL97}. To illustrate the idea, 
consider the Hill equation: 
\beq
\ddot{y} + [A - 2q P(2t)]y = 0
\l{eq:hill}
\eeq
where $P(2t)$ is any periodic function of the independent variable $t$. 
When $P(\cdot)  \propto \cos(\cdot)$, we have the Mathieu 
equation,  while $P(\cdot) = \mbox{cn}^2(\cdot)$ yields the Lam\'e 
equation, where  $\mbox{cn}(\cdot)$ is the elliptic cosine function. 
Now Eq. (\ref{eq:hill}) is equivalent to the one-dimensional
Schr\"odinger operator-eigenvalue problem:
\beq
{\cal L}(y) \equiv -\frac{d^2 y}{dx^2} + Q(x) y = \lambda y
\l{eq:shrod}
\eeq
under the transformations $x \arr t$, $Q(x) \arr 2q P(2t)$,  $\lambda 
\arr A$. 
We denote the spectrum of ${\cal L}$ by  $\sigma(\L)$, its 
complement by $\overline{\s(\L)}$.  The crucial point to 
notice is that the set of modes with positive Floquet index of Eq. 
(\ref{eq:hill}) contains  the set  $\overline{\s(\L)}$ of 
Eq. (\ref{eq:shrod}). By  exploiting this  equivalence  we will show that:

I -  Generically for almost-periodic potentials $Q(x)$, the corresponding
spectrum is a fractal Cantor set. This implies a Floquet index
which is positive on a dense subset of momenta $k \in [0,\infty)$.
As the strength of the coupling is varied, drastic changes to the 
spectrum can occur that are absent in the purely periodic case.

II - Preheating exists in stochastic inflation and the Floquet
index is positive with Lebesgue measure 1 for all momenta $k$. Estimates for 
$\mu_k$ are given in the perturbative and non-perturbative limits by 
Eq.'s (\ref{eq:shosmall}) and (\ref{eq:shobig}) respectively. 

These results describe geometrically the breakup of the 
stability-instability chart that in the case of the Hill equation 
(\ref{eq:hill}) has a neat band-structure. Visually, this is similar to 
the breakup of invariant tori in the KAM theory of chaos  \cite{chir79}. 
Similar results to those of II  above have recently been found independently 
\cite{HS97,zanc97} in specific cases for  the noise. In particular, 
Zanchin {\em et al} \cite{zanc97} rewrite Eq. 
(\ref{eq:hill}) as a first order matrix equation and then use 
Furstenberg's theorem \cite{furs63} regarding products of 
independent, identically distributed random matrices to obtain 
estimates of  $\mu_k$, showing that noise  increases the $\mu_k$  
over the periodic case and that $\mu_k > 0$ for 
all wavelengths, thus also demonstrating
the break-up of the stability bands, although in the 
context of small noise. 

The results of I and II above enlarge the known reheating territory (see 
Fig. 1).  The classical theory of reheating  was developed in 
the early 80's  \cite{80s}, with recognition of the importance of 
perturbative resonances \cite{res}. In this
case the resonance bands essentially dwindle  into a discrete 
sequence of lines. The paradigm shift of preheating \cite{KLS94},
showed that the resonance bands could be very broad at
large $q$, a fact that extends to 
gravitational waves \cite{BL97,bass97}. This was followed by 
approximations to  deal with the  non-perturbative effects 
\cite{Yosh95,Boyan},
and the idea of  non-thermal symmetry restoration induced by the  large 
quantum  fluctuations \cite{KLS95}. Recently stochastic resonance 
\cite{KLS97,GKLS97} has represented 
a shift away from  the simple picture of static resonance bands,  
due to the large  phase fluctuations  which, mod $2\pi$, behave randomly  
in the non-perturbative regime  in an expanding universe. Finally, the 
negative-coupling instability \cite{GPR97} represents the 
exciting possibility that modes can evolve where the spectrum of the 
corresponding Schr\"odinger operator is almost empty \footnote{This 
can be restated as the fact that the integrated density of states, 
proportional to  the rotation number \cite{moser1}, vanishes.} , and 
Floquet  indices are very large. This mechanism has been realised 
within the  framework of non-minimal fields coupled only to gravity, 
which therefore offers a  geometric  reheating  channel \cite{BL97} 
due to the oscillation of the Ricci curvature. 

\begin{figure}
\epsfxsize=3.2in
\epsffile{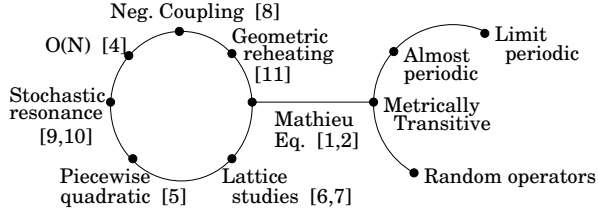}
\caption{A schematic map of  models and approximations in preheating. Minimal
references for techniques in the exactly periodic case are shown in 
brackets.  The right-hand branch corresponds to paradigms developed in 
this paper.} 
\l{fig:map} 
\end{figure}

In this paper we consider almost-periodic (section \ref{sec:cantor}) 
and random (section \ref{sec:random}) potentials which  are  
both  realizations of metrically transitive (i.e. ergodic and homogeneous)  
operators \cite{PF91,GS}.

\section{Cantor reheating}\label{sec:cantor}

We limit our discussion in this section to Minkowski 
spacetime \footnote{The general expanding FLRW case is 
contained within the spectral theory of Sturm-Liouville 
operators.}. Consider the inflaton, $\phi$,
and the two minimally coupled scalar fields  $\varphi$,  $\chi$ with the 
natural mass hierarchy $m_{\phi} \gg m_{\varphi} \gg m_{\chi}$. 
In this section we consider the effective  potential: 
\bea
V(\phi,\varphi,\chi) &=& \frac{m_{\phi}^2}{2} \phi^2  +  
\frac{\lambda}{4} \phi^4 +  \frac{m_{\varphi}^2}{2}\varphi^2  + 
\frac{m_{\chi}^2}{2}\chi^2 \\ \nonumber 
&+&  \frac{g^2}{2} F(\phi)\chi^2 
+ \frac{h^2}{2} J(\varphi)\chi^2
\l{eq:effpot}
\eea
$F(\cdot), J(\cdot)$ are assumed to be analytic in their respective 
arguments. The evolution of the $\chi_k$ modes is then \cite{KLS94}:
\beq
\ddot{\chi}_k + \left(\frac{k^2}{a^2} + m_{\chi}^2 
+ g^2 F(\phi) + h^2 J(\varphi) \right)\chi_k = 0
\l{eq:firsteom}
\eeq
Instead the inflaton zero-mode evolves according to:
\beq
\ddot{\phi} + m^2_{\phi, eff}\phi + \lambda \phi^3  
= 0 
\l{eq:zeromode}
\eeq 
where the frequency of oscillation is partially controlled by the 
effective mass:  
\beq
m^2_{\phi, eff} = m^2_{\phi} +  g^2 \frac{F'}{\phi} \langle \chi^2 
\rangle + 3 \lambda \langle \delta \phi^2 \rangle
\l{eq:effmass}
\eeq
In the case that $\langle \chi^2 \rangle = \langle \delta \phi^2 
\rangle = 0$, the inflaton simply oscillates with  constant period.    
However, when there are multiple-fields, or the mass acquires 
corrections due to quantum fluctuations, the period is no-longer 
constant, but may increase monotonically, oscillate or exhibit random 
fluctuations. As an example,  in the special case of a 
Yukawa interaction, $F(\phi) = \phi$, Eq. (\ref{eq:zeromode}) has a 
pure driving term $\propto g^2 \langle \chi^2
\rangle$. 

Understanding the behaviour of the Floquet indices in these 
more general cases is extremely complex and we therefore search for 
special, solvable classes,  which  naturally leads to the study of 
almost-periodic potentials  \cite{AS81}. In this  case,  the spectral 
theory becomes significantly  richer and more beautiful. 

In general, for a separable Hilbert space, as we have here, the 
spectrum can be decomposed w.r.t. an abstract measure $d\mu$ as
$\s = \s_{AC} \cup \s_{P} \cup \s_{SC}$ \cite{PF91},
where $\s_{AC}$ is the absolutely continuous, $\s_{P}$ is the pure point
and $\s_{SC}$ the singular continuous part of the spectrum.
An important constraint  is that the Floquet index must
vanish for $\lambda \in \s_{AC}$, i.e.   $\s_{AC} = \{\lambda \in {\bf
R}~|~ \mu_{\lambda} = 0\}$ bar a set of measure zero \cite{PF91}, and 
hence $\s_{AC}$ corresponds to modes with bounded evolution (extended
eigenfunctions).  On the complement,  $\overline{\s_{AC}}$, the Floquet
indices are  known to be positive, and hence in the case when $\s_{AC}$
is empty, $\mu_k > 0$ almost everywhere \cite{kirsch}.

To understand this better, consider the periodic  potential. In this case,
$\sp = \s_{AC}$ and on $\csp$, $\mu_{\lambda}
> 0$. It is the fact that $\s_{AC}$ is so large that forces the strong
band structure of the  Mathieu and Lam\'e equations. In the
Lam\'e  case, $\csp$ consists  of a {\em single} band and there are 
very few exponentially growing modes \cite{Boyan}. A comparison 
between the archetypal Cantor set and the spectrum of the Mathieu equation
is shown in Fig. (\ref{fig:cantorspec}).

The almost-periodic potentials are those for which the Fourier 
transform  consists of a frequency basis 
$\{\omega_i\}$ where the smallest vector space   
containing this basis, ${\cal  M}$,  is dense  in 
${\bf R}$. If ${\cal M}$ is generated  by 
finitely many $\omega_i$, then the potential, $Q(t)$, is quasi-periodic, 
i.e. $Q(t) =  f(\omega_1 t, ..., 
\omega_n t)$ with $f(t_1 + m_1, ..., t_n + m_n) = f(t_1, ..., 
t_n)$ with $m_i \in {\bf Z}$ and the $\omega_i$ pairwise 
incommensurate \cite{AS81}.

As a simple example consider Eq. (\ref{eq:firsteom}) 
with $F(\phi) = \phi^2~,~  J(\varphi) = \varphi^2$ and $\lambda = 0$.  
Then the equations for the quantum fluctuations of $\chi_k$  are: 
\beq
\ddot{\chi}_k + \left(\frac{k^2}{a^2} +  m_{\chi}^2 + 
g^2 \phi^2 +  h^2\varphi^2\right)\chi_k = 0 \,.
\l{eq:example}
\eeq
with $\phi \sim \sin(m_{\phi} t)$ and $\varphi \sim \sin(m_{\varphi} t)$.
When $m_{\phi}/m_{\varphi}$ is irrational, the  potential is 
quasi-periodic.  The spectrum of Eq. (\ref{eq:shrod}) for almost-periodic 
potentials  can be pure point and is, in a non-rigorous way, generally a 
nowhere  dense Cantor set \cite{sim82}. Hence, in the example above,  
for infinitely  many irrational  values of $m_{\phi}/m_{\varphi}$, we may 
expect that the spectrum of $\chi_k$ will  be  nowhere dense,  and 
consequently that almost all modes will grow exponentially.

However, unlike the random potentials to be 
considered in section (\ref{sec:random}), the Lebesgue measure of the 
spectrum, even if a Cantor set, need not be zero \cite{AS81} and $\s_{AC}$ 
may be non-empty. Indeed it is possible to have $\s = \s_{AC}$ 
\cite{moser2}. 

\begin{figure}
\epsffile{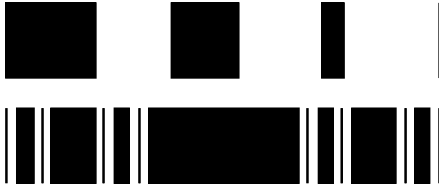}
\caption{A schematic diagram comparing the  spectrum of the Mathieu
equation (upper)  and that of a generic limit-periodic operator with 
fractal Cantor spectrum (lower). The black shading represents the 
complement of the spectrum, i.e. modes with $\mu_k \not = 0$. The 
Cantor spectrum is only
shown at second level due to resolution limitations. The same pattern
repeats itself in each white gap and hence the number of stable modes is
much smaller than in the periodic case.}
\l{fig:cantorspec} \end{figure}


More rigorous results exist in the case  that $Q$ is limit-periodic, 
i.e. it is a uniform limit of periodic potentials. A typical example is 
provided by:
\beq
Q(t) = \sum_{n=1}^{\infty} a_n \cos\left(\frac{2\pi 
t}{2^n}\right)~;~~\sum |a_n| < \infty
\l{eq:limit}
\eeq
For these potentials we can call on rigorous  theorems:

{\em \underline{Theorem 1} \cite{moser2} -- $\s(\L)$ of Eq. 
(\ref{eq:shrod}) is generically  \footnote{The space of limit periodic 
potentials is a complete metric  space and hence {\em generic} means here 
``for a dense $G_{\delta}$" e.g. \cite{sim82}.} a nowhere dense 
Cantor set for $Q$ an element of  the space  of limit periodic 
potentials. Hence  $\overline{\s({\L})}$ is dense in ${\bf R}$.} 

Thus the set of $k$ for which  $\mu_k \neq 0 $, is dense in ${\bf  R}^+$ 
despite the fact that $\s_{AC}$ is not empty. This existence of only a 
Cantor set of stable modes leads us to call this cantor reheating. An 
associated issue is what happens as the coupling to the potential
(analogous to $q$ in Eq. (1)) is increased. While in Eq. (1) this simply 
changes the breadth of the instability bands and magnitude of $\mu_k$ 
while always leaving $\s = \s_{AC}$, in the almost-periodic case this is 
not the case. Indeed, as in the example below,  the nature of the spectrum 
(w.r.t. the splitting  into $\s_P, \s_{SC}, \s_{AC}$) can change 
suddenly with $q$.

\subsection{The discrete almost-Mathieu equation}
 
Consider the discretized  Schr\"odinger eq. (\ref{eq:shrod}), which is 
also used (in higher dimensions) in numerical preheating studies  
\cite{KT96,PR97}. A special case is the almost-Mathieu equation with 
exhibits a rich variety of effects \footnote{For example, Eq. 
(\ref{eq:discrete}) also exhibits  the beautiful property of 
duality, similar to the S-duality of string theory, 
since under Fourier transform $q \rightarrow 1/q$ and $\lambda \rightarrow 
-\lambda/q$ \cite{AA80}.}: 
\beq
-y(x + 1) - y(x-1) + 2 q \cos(\alpha x + \omega) y(x) = \lambda y(x)
\l{eq:discrete}
\eeq
When $\alpha$ is rational the potential is periodic and  one has  
the band spectrum of the Mathieu equation. However, for irrational 
$\alpha$, the  spectrum is a Cantor set  generically for pairs  
$(q,\alpha) \in {\bf 
R}^2$  \cite{BS82}. This shows that the nowhere dense nature of the 
spectrum is not lost as one increases $q$ and hence moves from perturbative
reheating to broad-resonance preheating. Further, when  $\alpha$ is 
irrational,  the Floquet index has the lower bound \cite{AA80}:
\beq
\mu_k(q) \ge \ln |q|
\l{eq:lower}
\eeq
so that for $|q| > 1$, $\mu_k > 0$ and hence the absolutely continuous 
part of the spectrum, $\s_{AC}$, becomes empty. In this case, if $\alpha$ 
is a Liouville number \footnote{A Liouville number, $\alpha$, is irrational 
but  well approximated by rationals so that there exist integers $p_n,q_n 
\rightarrow  \infty$ and a number $C$ with  $|\alpha - p_n/q_n| \leq C 
n^{-q_n}$. }, then the spectrum is purely singular 
continuous, $\s = \s_{SC}$ \cite{AS82b}. Conversely if $|q| < 1$, the 
point part of the spectrum is absent for irrational $\alpha$ \cite{dely87}.

\section{Stochastic inflationary reheating}\label{sec:random}

We now consider the case where the potential $Q$ in Eq. 
(\ref{eq:shrod}) is random. This models for example,
the classical limit of stochastic inflation, where the 
dynamics of the local order parameter in a FLRW background,  
are described by \cite{mata97}:
\begin{equation}
\ddot{\phi} + 3H \dot{\phi} +  
V'(\phi)=\frac{H^2}{8\pi^3}V'''(\phi) \xi(t)\,, 
\l{eq:phi}
\end{equation}
where $H \equiv \dot{a}/a$ is the stochasticaly evolving Hubble constant 
\cite{CCV97},  $\xi$ is a coloured gaussian  
noise of unit amplitude with a correlation time of order $H^{-1}$, so 
that  the inflaton evolves stochastically. The origin of the noise is the
backreaction of quantum fluctuations with wavelengths shorter than the
coarse-graining scale \cite{MAB97,CCV97,HS97}. 

Here we will consider the potential $V(\phi) = \lambda \phi^4/4$ and 
as before an interaction term $g^2 \phi^2 \chi^2/2$. The quantum 
fluctuations of the fields $\phi$ and $\chi$  are
then given by \footnote{Here we neglect the backreaction of the 
produced field $\chi$ on the expansion; valid during the first phase of 
reheating before backreaction terminates the resonance.}: 
\beq 
(a^{3/2} \d\p_k \ddot{)}  + \left(\frac{k^2}{a^2} + 
3 \lambda \p^2 + \frac{3}{4} p - \frac{3 \lambda 
H^2}{4\pi^3} \xi(t)\right)(a^{3/2} \d\p_k) = 0\,, \l{eq:delphi}
\eeq
\beq
(a^{3/2} \chi_k\ddot{)}
+ \left(\frac{k^2}{a^2} + m_{\chi}^2 + \frac{3}{4} p + 
g^2 \p^2 \right)(a^{3/2} \chi_k) = 0 \,.
\l{eq:chi}
\eeq
where $p = \kappa(\ts{1\over2}\dot{\phi}^2 - V(\phi))$ is the pressure 
and  $\kappa = 8\pi G$. These equations are again equivalent to Eq. 
(\ref{eq:shrod}), but this
time with stochastic potentials, which are the opposite
extreme to periodic potentials since they generically have empty
$\s_{AC}$. In fact we have the following theorem:

{\em {\underline{Theorem 2}} \cite{kota84,simo83} -- If $Q(x)$ is a 
sufficiently random potential for Eq.  (\ref{eq:shrod}), then 
$\mu_{\lam} > 0$  for almost  all $\lam \in  {\bf R}$ and $\s_{AC}$ of 
$\L$ is empty with  probability one.}

Indeed we see that a positive Floquet exponent is
guaranteed for all modes bar a set of measure zero, and again reheating
is significantly different from the periodic case. Here ``sufficiently 
random"  means nondeterministic \cite{PF91} and is typically a 
requirement that correlations decay sufficiently rapidly. An example 
is a Gaussian random field whose correlation function has 
compact support. To proceed to obtain quantitative
estimates of the Floquet indices,  we exploit the fact that Eq.s 
(\ref{eq:delphi},\ref{eq:chi})  have the form of stochastic 
harmonic oscillators.

\subsection{Explicit estimates for the Floquet indices}

Consider the stochastic harmonic oscillator, with frequency given by
$\omega^2 = \kappa^2 + q \xi(t)$, where $q$ is a 
dimensionless coupling to the mean-zero coloured stochastic process 
$\xi(t)$ and 
$\kappa^2 = k^2/a^2$ in our case. The Floquet index has been shown to be 
strictly positive for all $q$  \cite{Arn84}. It has also been  
related to the  spectral density of fluctuations via  averaging over 
the second moments \cite{vK76} of the random process $\xi$ \cite{CGR93}:
\beq 
\mu_k = \frac{\int_0^{\infty} \langle\xi(t)\xi(t - t')\rangle \cos[2
\langle \omega^2 \rangle t'] dt'}{2|\langle\omega^2\rangle|}\,.
\l{eq:kubo}
\eeq
This has the form of a fluctuation-dissipation theorem since 
fluctuations in the inflaton field determine the dissipation rate into 
other  fields. In the case that $\xi(t)$ is a mean zero ergodic 
Markov process, $\mu_k$ can be explicitly estimated as \cite{papa}: 
\beq
\mu_k = \frac{\pi}{4} \frac{q^2}{\kappa^2} \hat{f}(2\kappa)  + 
O(q^3) 
\l{eq:shosmall}
\eeq
where $\hat{f}$ is the
Fourier transform of the expectation value of the two-time correlation 
function $\langle \xi(t)\xi(t - t')\rangle$.  Note that $\mu_k \propto 
k^{-2}$, so that although all  modes grow exponentially, the Floquet 
index is, as in the periodic  and stochastic resonance \cite{KLS97} 
cases, a rapidly decreasing  function of $k$. In  the broad-resonance 
limit, $q \rightarrow  \infty$, we write  $\kappa^2 = \kappa^2_0 + 
q \kappa^2_1$, which gives (assuming $max~ \xi < \kappa^2_1$) 
\cite{papa}: 
\bea
\mu_k &=& \frac{\kappa_1}{4\pi} \int_0^{2\pi} d\theta \int d\xi 
\frac{\sqrt{\kappa^2_1 - \xi}}{\kappa^2_1 - \xi \cos^2 \theta}  
G(\ln(\kappa^2_1 - \xi \cos^2 \theta)) \nonumber\\  &+& O(1/\sqrt{q}) 
\l{eq:shobig}
\eea 
where $G$ is the infinitesimal generator of $\xi(t)$  
defined by the  limit of the operator sequence:
$G = lim_{t\rightarrow 0} (U_t - {\bf I})/t$. Here ${\bf I}$ is the 
identity operator and $U_t$ is a family of
operators on the space of bounded continuous functions $f$, defined by 
$U_t f(x) = E[f(\xi(t))|\xi(0) = x]$, 
where $E$ denotes the  expectation operator \cite{GS}. 
Again $\mu_k > 0$ 
for all $k$, and given a model of $\phi$ evolution, one can explicitly 
estimate $\mu_k$, and  hence the numbers of produced particles $n_k$  
and variances $\langle \delta\phi^2 \rangle$ and $\langle \chi^2 \rangle$.  

\section{Amplification of gravitational waves}

Here it is demonstrated that the evolution equatons for gravitons can 
be cast in the form of the Schr\"odinger Eq. (\ref{eq:shrod}). Within the 
Bardeen formalism  \cite{BL97},  or by using the  Weyl tensor,  there 
exists a strong correspondence between  gravitational waves and  scalar 
fields  during reheating \cite{bass97}. In the Bardeen approach, 
gravitational waves are described by transverse-traceless metric 
perturbations, with mode functions $h_k$ satisfying \cite{BL97}: 
\beq
(a^{3/2} h_k)\ddot{} + \left(\frac{k^2}{a^2} + \frac{3}{4}p 
\right)(a^{3/2} h_k) = 0\,, \l{eq:bard2} \eeq 
so that $Q \arr  - 3p/4$,  $k^2/a^2 \arr \lambda$ and $a^{3/2} h_k \arr 
y$  establishes 
the equivalence with Eq. (\ref{eq:shrod}). This  implies that during 
reheating with an almost-periodic or stochastic 
inflaton evolution, all wavelengths will be amplified, rather than just 
those in the Floquet  instability bands.  In the stochastic case with 
small noise, Eq. (\ref{eq:shosmall}) shows that 
the tensor spectrum will acquire a tilting with $\mu_k \sim k^{-2}$, 
causing further subtleties for the inflationary potential  
reconstruction program.

\section{Concluding remarks}

Using spectral theory we have shown the  breakup of the stability 
bands of the Mathieu equation. For constant $q$, the set of stable modes  
becomes a Cantor set generically for  almost-periodic 
inflaton evolution  and a set of measure zero if the evolution is 
sufficiently random. This implies that quantum corrections to the 
effective mass will lead to an exponential  growth of all modes of the 
reheated field, and of the  gravitational wave background produced during 
inflation. Further, there exists the  possibility, absent in the periodic 
case, of drastic qualitative changes to the nature of the spectrum at 
certain transitional values of $q$, as demonstrated by the discrete 
almost-Mathieu equation. 
   
This is of relevance to the issue of non-thermal symmetry restoration,
since the quantity $\int_{\ov{\s(\L)}} \mu_k dk$ is much 
larger in the classes considered here. This should reflect directly in 
the number  of particles produced, $n_k$, the non-thermality of the 
spectrum  and the variances of the fields.

We have largely limited ourselves here to the study of spectral changes 
as the potential is changed and found drastic changes from the simple 
band structure of Floquet theory. However, as a 
warning,  beautiful counter-examples exist, though they are suitably 
rare. Firstly, using inverse scattering theory for Eq. (\ref{eq:shrod}), 
one can show that there exist, and infact derive all examples of,
almost-periodic  potentials which have band spectra similar to the 
periodic case. An example is given by the Bargmann potentials 
\cite{moser1}. Hence it is not always true that the spectrum of an 
almost-periodic operator must be a Cantor set.

Neither is it true that altering the coupling to the potential always
induces a change in the spectrum. From  the  theory  of isospectral 
deformations of the Schr\"odinger equation it is known that if $Q(x,y)$ is a 
potential of Eq. (\ref{eq:shrod}) and satisfies the KdV equation $Q_y - 
6Q Q_{x} + Q_{xxx} = 0$,  then $\sp$ is independent of $y$ and there thus 
exist 1-parameter families of  potentials with the same spectra 
\cite{moser1}. The connections with  integrable  systems become clear, 
and hence so do the rarity of such events.

\section{Acknowledgements}

This paper is dedicated to the late Lando Caiani, who shared many of his 
insights with me. I also thank Andrew  Matacz, David Kaiser, Davide 
Guzzetti,  Finlay Thompson and Stefano Liberati for enlightening discussions.

\end{document}